\documentclass[prl, twocolumn, superscriptaddress, showpacs]{revtex4}
\usepackage{graphicx, subfigure}
\usepackage{dcolumn}
\usepackage{bm}
\usepackage{multirow}
\usepackage{color}

\begin{document}

\title{All-Electron Path Integral Monte Carlo Simulations of Warm Dense Matter: Application to Water and Carbon Plasmas}

\author{K. P. Driver}

\email{kdriver@berkeley.edu}

\homepage{http://militzer.berkeley.edu/~driver/}

\affiliation{Department of Earth and Planetary Science, University of California, Berkeley, California 94720, USA}

\author{B. Militzer}

\affiliation{Department of Earth and Planetary Science, University of California, Berkeley, California 94720, USA}

\affiliation{Department of Astronomy, University of California, Berkeley, California 94720, USA}

\date{\today}

\begin{abstract}{ 
We develop an all-electron path integral Monte Carlo (PIMC) method
with free-particle nodes for warm dense matter and apply it to water
and carbon plasmas. We thereby extend PIMC studies beyond hydrogen and
helium to elements with core electrons. PIMC pressures, internal
energies, and pair-correlation functions compare well with density
functional theory molecular dynamics (DFT-MD) at temperatures of
(2.5-7.5)$\times10^5$ K and both methods together form a coherent
equation of state (EOS) over a density-temperature range of 3--12
g/cm$^3$ and 10$^4$--10$^9$ K.}\end{abstract}


\maketitle

The development of first-principles methodology for warm, dense matter
(WDM) is one of the great challenges of modern materials theory. A
need for rigorous simulation of WDM has escalated with intensified
interest in advanced energy technologies~\cite{Cook2006}, physics and
chemistry of solar and extrasolar planets~\cite{Fortney2009}, shock
compressed matter~\cite{Koenig2005}, and different types of
plasma-matter interactions~\cite{Drake2006}. The standard
first-principles method, Kohn-Sham density functional theory molecular
dynamics~\cite{Tse2002} (DFT-MD), produces accurate equations of state
in the lower temperature range of the WDM regime. The maximum
accessible temperature is limited, however, because the number of
partially occupied orbitals eventually becomes computationally
intractable~\cite{Surh2000}. On the other hand, the many-body path
integral Monte Carlo (PIMC) method~\cite{Ceperley1995} is naturally
formulated to study high temperature dependence of materials. Ideally,
PIMC and DFT together can produce a coherent equation of state for the
entire WDM regime and cross-validate each other at commonly accessible
temperatures. However, PIMC has not yet been developed to study
systems with core electrons. Indeed, PIMC studies up to now have been
limited to plasma states of
hydrogen~\cite{Pierleoni1994,Militzer2000,Militzer2001a} and
helium~\cite{Militzer2006,Militzer2009}. In this letter, we develop an
all-electron PIMC method for first-row elements and combine results
with DFT-MD to produce comprehensive equations of state for water and
carbon in the WDM regime for a density-temperature range of 3--12
g/cm$^3$ and 10$^4$--10$^9$ K.

\begin{figure}[t]
  \begin{center}
        \includegraphics[width=8.6cm]{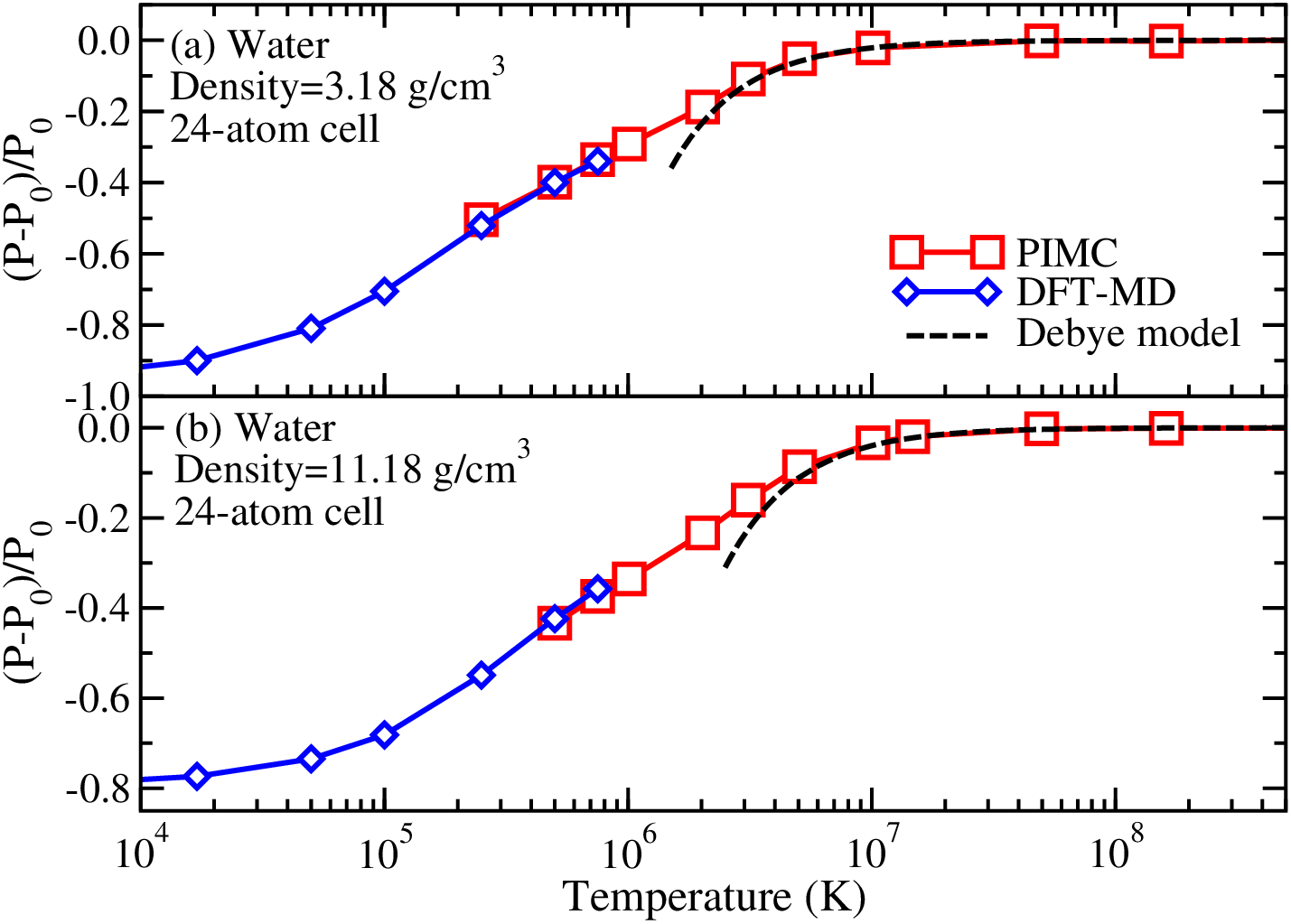}
  \end{center}

    \caption{(color online) Comparison of excess pressure relative to
      the ideal Fermi gas plotted as a function of temperature for
      water.}

  \label{fig:PvsTWater}
\end{figure}

The central characteristic of a material in the WDM regime is that the
electron-ion interaction becomes comparable to the electron kinetic
energy and, therefore, effects of bonding, ionization,
exchange-correlation, and quantum degeneracy are all important. The
analytic methods of condensed matter and plasma
physics~\cite{Ebeling1976} are typically not reliable without
experimental input. One must turn to the the numerical,
first-principles PIMC and DFT-MD methods which accurately capture the
many-body physics in the WDM regime without empirical parameters or
corrections. However, first-principles methods utilize certain
approximations and one must compare with experimental data if
available.

The key approximation in DFT is that of the exchange-correlation
potential, which describes all the many body interactions. The
exchange-correlation potentials used in nearly all condensed matter
calculations are constructed from zero temperature quantum Monte Carlo
calculations of the electron gas~\cite{Ceperley1980}. In the WDM
regime, temperatures are at or above the Fermi temperature and
electrons are excited relative to their ground state. Therefore,
without further validation, the exchange-correlation potential cannot
be assumed to provide an accurate description in the WDM regime. 

In DFT calculations it is also common to replace the core electrons in
each atom with a pseudopotential.  Typically, highest accuracy is
obtained with a non-local pseudopotential which depends on the energy
and angular momentum components in core states.  However, in the WDM
regime, it is possible to excite electrons out of core levels. The
pseudopotential approximation may break down and should always be
compared with all-electron calculations. Additionally,
finite-temperature DFT uses a Fermi-Dirac function to allow for
thermal occupation of single-particle electronic
states~\cite{Mermin1965}, but requires an increasing number of bands
with temperature, crippling its efficiency at extreme
temperatures. Orbital-free density functional methods aim to overcome
such thermal band limitations, but several challenges remain to be
solved~\cite{Lambert2006}.

The PIMC method avoids the band structure calculation and
exchange-correlation approximation by being directly defined from the
path integral formulation of quantum statistics. PIMC stochastically
solves the full finite-temperature quantum many-body problem by
treating electrons and nuclei equally as paths and addresses all of
WDM characteristics on an equal footing. All finite-temperature
properties of a material are then readily calculated from the thermal
density matrix. In contrast to DFT, PIMC efficiency increases with
increasing temperature as particles become more classical and fewer
time slices are needed, scaling inversely with temperature. A
non-local pseudopotential formulation of PIMC does not yet
exist~\cite{Jabbour1994} and this is why PIMC calculations so far have
been limited to hydrogen and helium. PIMC calculations presented here
treat all electrons explicitly.

The only uncontrolled approximation in PIMC is that of the nodal
surface to deal with the fermion sign problem. Unchecked, the fermion
sign problem leads to a cancellation of positive and negative
contributions to the density matrix which causes large fluctuations in
computed averages. One solution to this problem is the so-called
fixed-node approximation in which the location of the nodes are fixed
to a known trial nodal structure in order to guarantee positive
contributions to the thermal density matrix. The form of the density
matrix does not enter the PIMC computation, only the location of the
nodes.

The PIMC method we present here employs a free-particle nodal
structure, which is expected to be accurate for systems that are
almost fully ionized. One could assume accurate calculations of
heavier elements requires very high temperatures where atomic cores
are ionized also. However, for hydrogen, PIMC with free-particle nodes
has provided reliable results at much lower temperatures where only
partial ionization occurs~\cite{Militzer2001a}. The PIMC results
presented for water and carbon here will demonstrate that accurate
pressures and energies are obtained for temperatures so low that the
1s states are fully occupied and the 2s states are partially
occupied. Analyses of the DFT band occupations show that as the 2s
states become less than 60\% occupied for $T \ge 2.5\times10^5$ K,
PIMC and DFT results agree. 

In order to explain this result, we first note that no nodes are
needed to describe an isolated, doubly occupied 1s state. Our results
for water and carbon indicate that free particle nodes also work in cases
where the 1s state is doubly occupied and all other electrons are
ionized. This may be because only one orbital out of many in the
Slater determinant is not characterized well. As the occupation of the
2s state exceeds 60\% at lower temperatures, the PIMC pressures and
energies become inaccurate because free particle nodes cannot yield
the correct shell structure around the nucleus~\cite{Shumway2006}. Our
results will show, for first-row elements, free particle nodes remain
sufficiently accurate at low enough temperatures to overlap with the
highest temperature DFT data. This allows the two methods to
cross-validate each other and form a single coherent equation of state
for all temperatures.

\begin{figure}[t]
  \begin{center}
        \includegraphics[width=8.6cm]{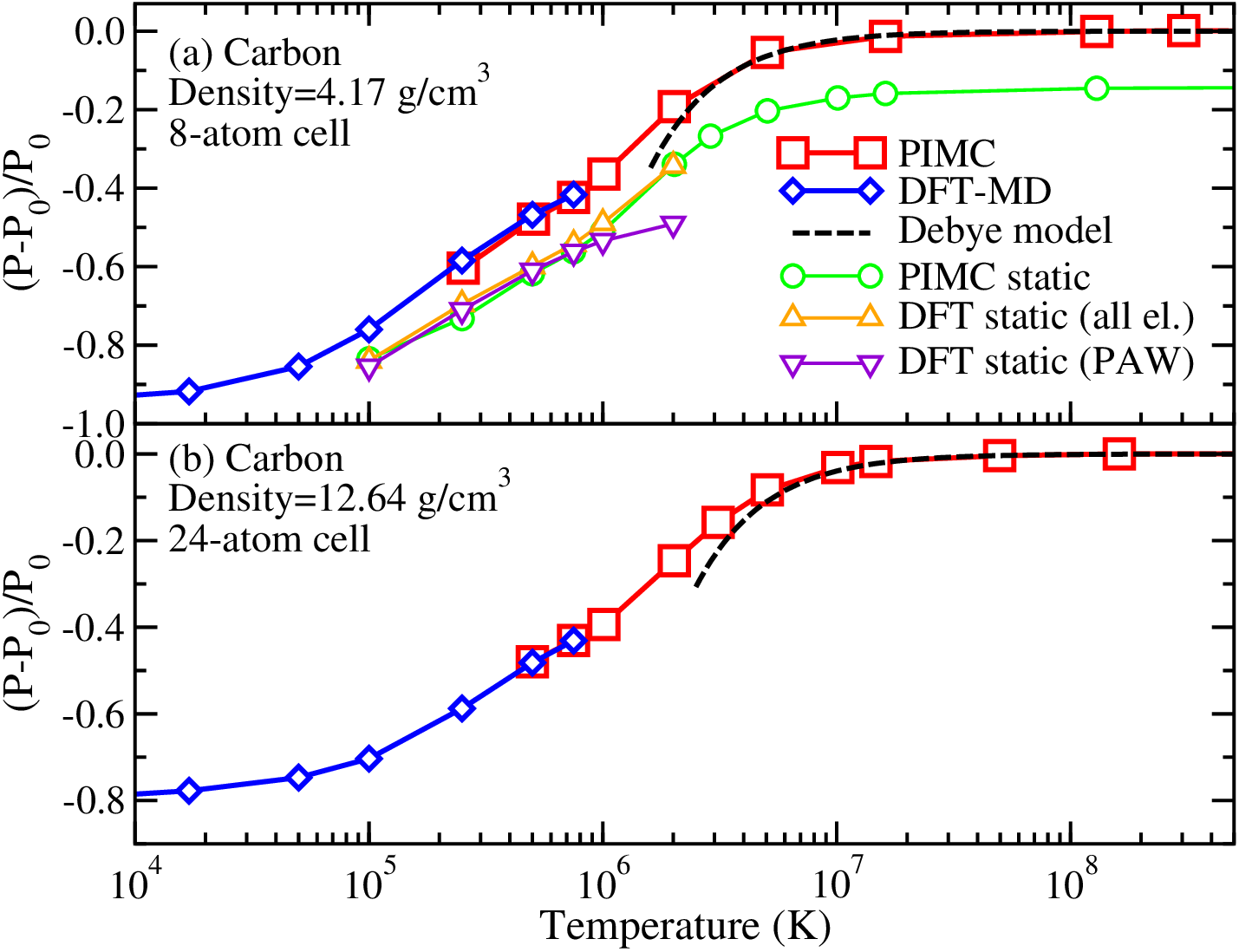}
  \end{center}

    \caption{(color online) Comparison of excess pressures relative to
      the ideal Fermi gas plotted as a function of temperature for
      carbon.}

  \label{fig:PvsTCarbon}
\end{figure}

As a first application to test our method, we study water because it
is one of the most prevalent materials in nature and knowledge of its
electronic properties in the WDM regime is crucial for understanding
aspects of astrophysical objects, such as the interiors of giant gas
planets. Reports suggest Uranus, Neptune, Jupiter, and Saturn contain
significant amounts of
water~\cite{Nettelmann2008,Redmer2011,Celliers2004}. In addition to
its rich solid and fluid phases, water is known for its superionic and
plasma phases as well as an insulator-to-metal transition at extreme
densities and temperatures. Recent DFT-MD
simulations~\cite{French2009} have computed the equation of state of
water up to 2$\times10^4$ K and 15 g/cm$^3$, improving upon the older
SESAME 7150~\cite{SESAME1992} table comprised of a number of analytic
models and MD using empirical potentials.

\begin{figure}[t]
  \begin{center}
        \includegraphics[width=8.6cm]{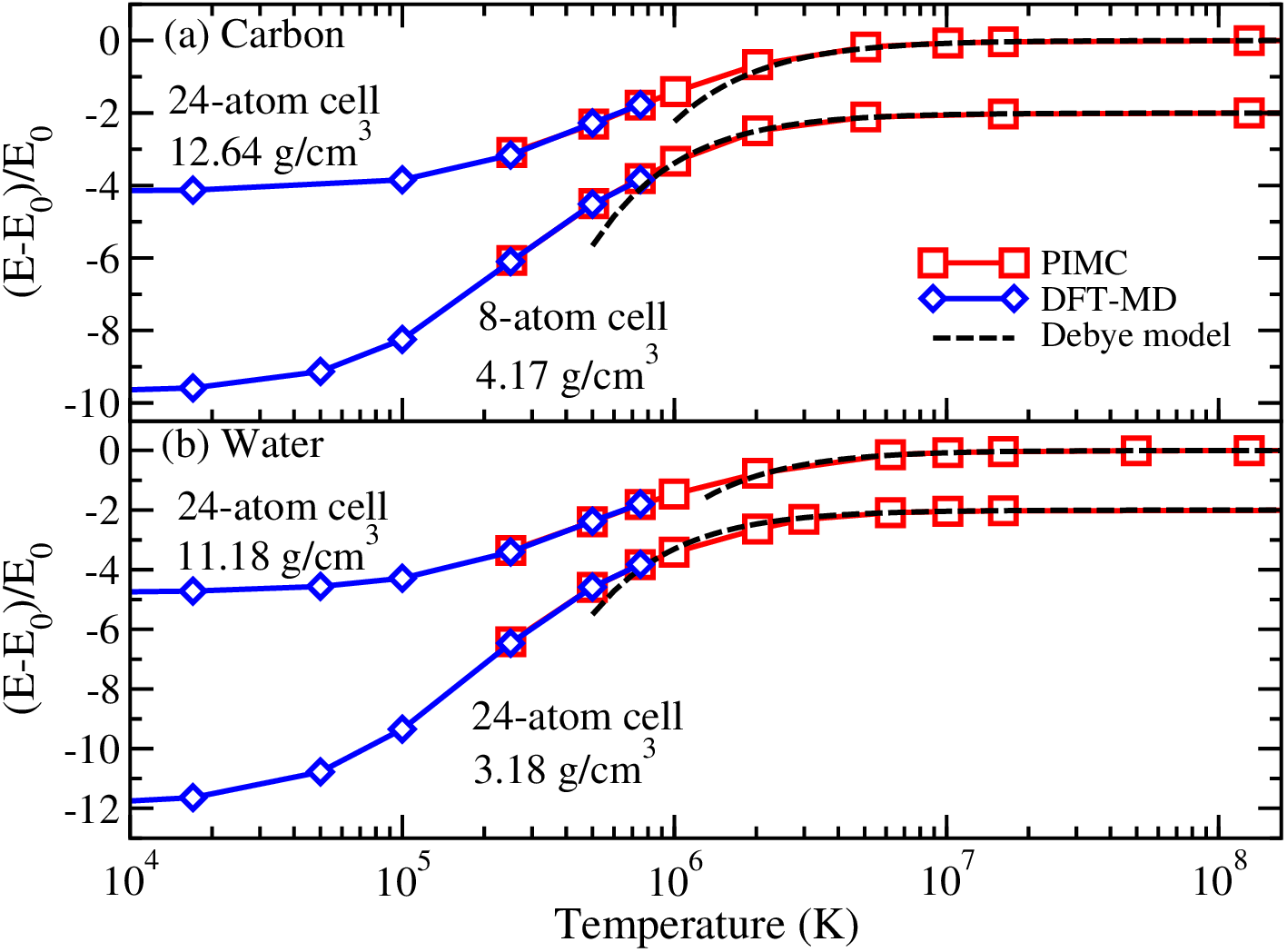}
  \end{center}

    \caption{(color online) Comparison of excess internal energies
      relative to the ideal Fermi gas plotted as a function of
      temperature for (a) carbon and (b) water. The lower density data
      has been shifted by a constant, -2, in both cases.}

  \label{fig:IEvsT}
\end{figure}

As a second application, we study carbon at high pressures and
temperatures for its importance in future energy technologies. In
inertial confined fusion experiments, carbon is used as an ablator for
target capsules. The performance of the ablator is heavily dependent
on the equation of state in the WDM regime~\cite{Hammel2006}. There
have been a number of attempts to construct carbon equations of state
in the WDM regime, including free energy
models~\cite{Kerley2001,Potekhin2005} and DFT-MD~\cite{Correa2008},
but they ultimately resort to more approximate Thomas-Fermi-based
models that cannot describe any chemical bond. 

For our PIMC simulations, the Coulomb interaction is incorporated via
pair density matrices derived from the eigenstates of the two-body
Coulomb problem. 
A sufficiently small time step is determined by
converging total energy as a function of time step until the energy
changes less than 0.2\%. For both water and carbon, we use a time step
of 0.0078125 Ha$^{-1}$ for temperatures below 5$\times10^6$ K and, for
higher temperatures, the time step decreases as $1/T$ while keeping at
least four time slices in the path integral. In order to minimize
finite size errors, the total energy is converged to better than 0.2\%
for a 24 atom simple cubic cell.

The DFT-MD simulations were performed with either the ABINIT
code~\cite{ABINIT} using all-electron pseudopotentials or with the
Vienna $ab~initio$ simulation package (VASP)~\cite{VASP} using the
projector augmented-wave method~\cite{PAW}. MD uses a NVT ensemble
regulated with a Nos\'{e}-Hoover thermostat. Exchange-correlation
effects are described using the Perdew-Burke-Ernzerhof generalized
gradient approximation~\cite{PBE}. Electronic wave functions are
expanded in a plane-wave basis with a energy cutoff of at least 1500
eV for water and at least 900 eV for carbon in order to converge total
energy to chemical accuracy.  Size convergence tests indicate that
total energies are converged to better than 0.2\% in a 24 atom simple
cubic cell.  Convergence of the number of orbitals to a prescribed
thermal occupation of less than 1$\times10^{-4}$ requires up to 1500
bands at 7.5$\times10^5$ K for a 24-atom cell.  All simulations are
performed at the gamma-point of the Brillouin zone, which converges
total energy to better than 0.1\% at relevant high temperatures.

\begin{figure}[t]
  \begin{center}
        \includegraphics[width=8.6cm]{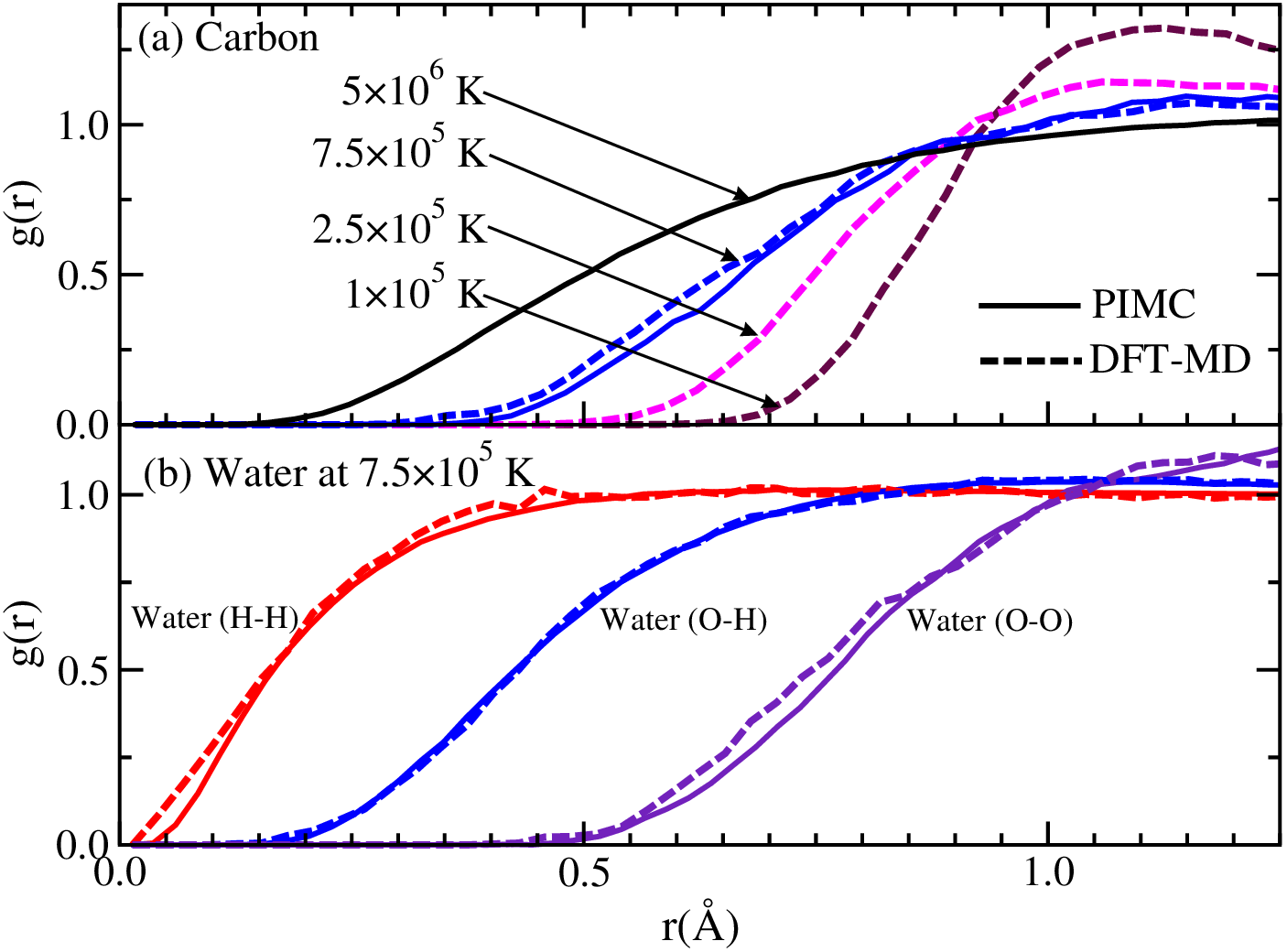}
  \end{center}

    \caption{(color online) Nuclear pair-correlation functions for
      (a) carbon and (b) water.}

  \label{fig:gofR}
\end{figure}

Figures 1 and 2 compare pressures obtained for water and carbon,
respectively, from PIMC, DFT-MD, and
Debye-H\"{u}ckel~\cite{DebyeHuckel} simulations. Water is studied at
fixed densities of 3.18 and 11.18 g/cm$^3$ and carbon is studied at
4.17 g/cm$^3$ and 12.64 g/cm$^3$. The two densities in each case
correspond to a pressure of 1 Mbar and 50 Mbar at zero temperature.
Pressures, P, are plotted relative to a fully ionized Fermi gas of
electrons and ions with pressure, P$_{0}$, in order to compare only
the pressure contributions that result only from particle
interactions. PIMC and DFT-MD results for $(P-P_0)/P_0$ agree to
better than 0.03 in the range of 2.5$\times10^5$ to 7.5$\times10^5$
K. Convergence tests show that results are equally well converge in
24-atom and 8-atom simulation cells. The excellent agreement allows
for cross-validation which implies the zero temperature DFT
exchange-correlation potential remains valid at high temperatures and
that the free-particle nodal approximation is valid in PIMC when atoms
are only partially ionized.  The two methods have comparable
computational cost in the overlap region, but DFT computational cost
starts to become prohibitive beyond 7.5$\times10^5$ K, and free
particle nodes break down below 2.5$\times10^5$ K. 

In addition, Fig. 2 compares the instantaneous pressures obtained for
a fixed configuration of carbon nuclei at various electronic
temperatures using PIMC, DFT with all electron
pseudopotentials, and DFT with VASP PAW pseudopotentials. Agreement
between PIMC and DFT with all electron pseudopotentials is very good
from 1$\times10^5$ to 2$\times10^6$ K. However, beyond 7.5$\times10^5$
K, PAW DFT no longer predicts the correct temperature dependence,
indicating that the missing contributions of core excitations to the
total energy become significant. All electron DFT is too
computationally demanding to perform calculations with moving nuclei.

In Fig. 3, the internal energies, E, are plotted relative to the ideal
internal energy, E$_{0}$. PIMC and DFT-MD results for $(E-E_0)/E_0$
agree to better than 0.04 in the range of 2.5--7.5$\times10^5$ K for
water and carbon. Convergence tests show that results are equally well
converged in 24-atom and 8-atom simulation cells. PIMC extends the
equations of state to the weakly interacting plasma limit at high
temperatures, in agreement with the Debye-H\"{u}ckel model. The DFT-MD
and PIMC methods together form a coherent equation of state over all
temperatures.

Figure 4 shows nuclear pair-correlation functions for carbon and water
using PIMC and DFT-MD. Fig. 4(a) demonstrates the sensitive
temperature dependence of structural properties for carbon. Water
pair-correlations are shown in Fig. 4(b) at a single temperature of
7.5$\times10^5$ K. Simulations use a 24-atom simulation cell size,
which converges pair-correlation curves to better than 10\%. The PIMC
and DFT pair correlation curves essentially lie on top of each other
with the maximum deviation being 17\% for carbon at r=0.63 \AA. The
results demonstrate that PIMC and DFT predict consistent structural
properties in addition to the equation of state.

In conclusion, we have developed an all-electron path integral Monte
Carlo method using free-particle nodes that allows for calculations of
materials composed of first-row elements and mixtures thereof. Our
computations of pressures, internal energies, and pair-correlation
functions for water and carbon demonstrate that PIMC and DFT can
cross-validate each other in a commonly accessible temperature range
and provide an accurate, coherent equation of state ranging from
ambient conditions to the plasma limit. The excellent agreement
between our PIMC method and DFT-MD validates the use of free-particle
nodes for partially-ionized first-row elements and the use of
zero-temperature exchange correlation functionals at high temperature.

\begin{acknowledgments}
  
  This research is supported in part by the UC Berkeley lab fee grant
  and by the NSF. Computational resources were provided by TAC, NCAR,
  and NERSC.

\end{acknowledgments}


\end{document}